# Infrared spectroscopy of radioactive hydrogen chloride H$^{36}$Cl

Santeri Larnimaa[a,]*, Markku Vainio[a,b], Ville Ulvila[c]

[a] Department of Chemistry, University of Helsinki, Helsinki, Finland

[b] Photonics Laboratory, Physics Unit, Tampere University, Tampere, Finland

[c] VTT Technical Research Centre of Finland Limited, Espoo, Finland

* Corresponding author

    Postal address: Department of Chemistry, University of Helsinki, P.O. Box 55, FI-00014 Helsinki, Finland

    E-mail address: santeri.larnimaa@helsinki.fi

## Abstract

We present the first report of optical absorption spectroscopy of H$^{36}$Cl, a radioactive isotopologue of hydrogen chloride. We used Fourier-transform infrared spectroscopy to determine the line center wavenumbers of the fundamental rovibrational band lines P(10)–R(10) and the first overtone band lines P(1)–R(7) with total uncertainty of less than 0.0018 cm$^{-1}$ (60 MHz) and 0.007 cm$^{-1}$ (0.2 GHz), respectively, at 68 % confidence level. We also performed a rotational analysis on the bands to determine the related molecular constants. We further compared the linewidths and relative intensities of the lines to those of the stable isotopologues H$^{35}$Cl and H$^{37}$Cl. The new spectroscopic information assists in developing optical instrumentation for the detection of H$^{36}$Cl.

## 1. Introduction

Hydrogen chloride is an important molecule that plays a role in atmospheric chemistry [1,2], astrochemistry [3-5], the semiconductor industry [6], and in the research of volcanic activity [7-10] and planetary atmospheres [11-13]. While the rovibrational bands of HCl were studied in detail already in the 1960s [14-16], early works on HCl date to the beginning of the 20[th] century to the development of early spectrometers and the acceptance of quantum mechanics [17-19]. In recent years, there have been many publications that report increasingly accurate line parameters [20-24]. The naturally occurring stable isotopologues H$^{35}$Cl, H$^{37}$Cl, and their deuterated forms are usually studied. However, to the best of our knowledge, there are no prior reports on optical spectroscopy of H$^{36}$Cl.

Chlorine-36 is a radioactive isotope of chlorine that is formed naturally in the atmosphere [25] and can be used, for example, for groundwater dating [26-29]. The species is also formed in nuclear facilities by neutron activation of $^{35}$Cl impurities found in reactor materials [30-33]. This is a major concern due to the long (ca. 301 000 years) half-life of $^{36}$Cl. Some of $^{36}$Cl may be released to the environment in the form of H$^{36}$Cl.

Owing to their high sensitivity and selectivity, optical spectroscopy methods are well suited for the detection of radioactive compounds in the gas phase [34]. These methods can also offer a more affordable and portable alternative to the current state-of-the-art method, Accelerator Mass Spectrometry [35]. In addition, spectroscopic methods do not depend on the specific activity of the species of interest, which might lead to less laborious sample preparation than in Liquid Scintillation Counting [36], as the need for a complex radiochemical separation is circumvented.

Prior research on optical spectroscopy of radioactive compounds has focused especially on laser spectroscopy of radiocarbon dioxide $^{14}$CO$_2$ [34,37-45]. One goal has been to develop highly sensitive and portable field instruments for emission monitoring of this species in nuclear facilities and their decommission sites. In general, the development of laser spectroscopy instrumentation for the detection of radioactive molecules is hindered by the lack of spectroscopic information. For example, we only recently reported the first absorption measurements and rotational analysis of radiocarbon methane $^{14}$CH$_4$ [46,47], another important species of concern in nuclear facility emission monitoring [48].

Here we report the results of our Fourier-transform infrared spectroscopy (FTIR) measurements of the fundamental and first overtone rovibrational bands of H$^{36}$Cl. These results aid the development of optical instrumentation for monitoring this species. In the following sections, we first describe the sample preparation, measurement and line-fitting procedures, after which we summarize the main results. Finally, we present the rotational analysis and discuss the results for the overtone band. Supplementary information contains information about the experimentally observed linewidths and relative intensities of the H$^{36}$Cl fundamental band lines, including a comparison to those of H$^{35}$Cl.

## 2. Sample preparation and measurement conditions

The gaseous H$^{36}$Cl sample was prepared by pipetting 50 µl of $^{36}$Cl-enriched NaCl solution (3.7 MBq/ml activity concentration; American Radiolabeled Chemicals, Inc.) into a 10-cm long IR Quartz cell (Suprasil 300; FireflySci Type 34). The aqueous sample was subsequently dried to solid NaCl using a flow of compressed air, after which a few drops of concentrated sulfuric acid were added and the cell was closed with PTFE stoppers. Finally, the reaction of the solid NaCl with sulfuric acid to produce HCl was initiated by tilting the sample cell to combine the reagents. This simple gas cell system was not equipped with a pressure gauge or a temperature meter to minimize the sample cell volume and sample losses.

Given the specific activity of $^{36}$Cl (1220 Bq/µg), the total amount of $^{36}$Cl in the sample was expected to be 4.2 µmol. Based on the FTIR measurements, we obtained 1.6 mbar of gaseous H$^{36}$Cl. This estimate assumes the HITRAN H$^{35}$Cl line intensities for H$^{36}$Cl (see Supplementary Note 4). This means that the yield into gas phase was 43 %, assuming 296 K temperature and 28 ml volume of the sample cell. The sample also contained approximately 0.4 mbar and 0.8 mbar of the stable isotopologues, H$^{35}$Cl and H$^{37}$Cl, respectively. The total partial pressure of HCl was 2.7 mbar. Since the pressure increase in the sample cell due to the produced HCl is small and the partial pressure of water in concentrated sulfuric acid is negligible [49], it is reasonable to assume 1 atm (1013.25 mbar) pressure in the cell during the measurements. The lab pressure was within a few mbar of this pressure upon filling the sample cell.

## 3. Measurement of the fundamental band

The H[36]Cl measurements in the mid-infrared (MIR) were executed using a commercial FTIR instrument (Bruker IFS 120HR). We used a Globar light source, Ge-on-KBr beam splitter, and liquid-nitrogen-cooled InSb detector. The raw spectrum is a result of processing 10 co-added double-sided interferograms using the Mertz method [50]. We used the manufacturer-specified resolution setting of 0.02 cm$^{-1}$ (600 MHz). The manufacturer defines the resolution as the full width at half maximum (FWHM) of the assumed instrument line shape (ILS) function if triangular apodization was used; we used Norton-Beer medium apodization [51], which corresponds to 0.0095 cm$^{-1}$ (285 MHz) half width at half maximum (HWHM) of the resulting ILS function. Table 1 lists other relevant FTIR instrument settings used.

Table 1. Relevant FTIR instrument settings used for the MIR measurements.

| Light source | Globar |
| --- | --- |
| Beam splitter | Ge-on-KBr |
| Detector | InSb |
| Aperture diameter | 1 mm |
| Apodization function | Norton-Beer medium |
| Assumed HWHM of ILS function | 0.0095 cm$^{-1}$ (285 MHz) |
| Number of co-added interferograms | 10 |
| Type of interferograms | Double-sided |
| Interferogram processing method | Mertz |
| Zero-filling factor | 2 |
| Optical band-pass filtering | 2614–3378 cm$^{-1}$ |
| Digital band-pass filtering | 10–20 kHz (1975–3950 cm$^{-1}$) |
| HeNe reference laser down-converted frequency | 80 kHz |

The 100 % transmission baseline was determined manually from the HCl spectrum by first low-pass filtering the spectrum, excluding the HCl peaks from the data, and finally by using Savitzky-Golay filtering [52] and defining the result as the baseline. The spectrum was then divided by this baseline to obtain the HCl transmission spectrum. The result was further converted into absorption spectrum using the Naperian Beer-Lambert law [53]

$$\alpha(\tilde{v}) = -\frac{1}{L}\ln\left(\frac{I(\tilde{v})}{I_0(\tilde{v})}\right),$$

where $\alpha(\tilde{v})$ is the absorption coefficient, $L$ is the absorption path length (10 cm), $I(\tilde{v})$ is the raw spectrum, and $I_0(\tilde{v})$ is the 100 % transmission baseline. The resulting absorption spectrum is shown in Fig. 1.

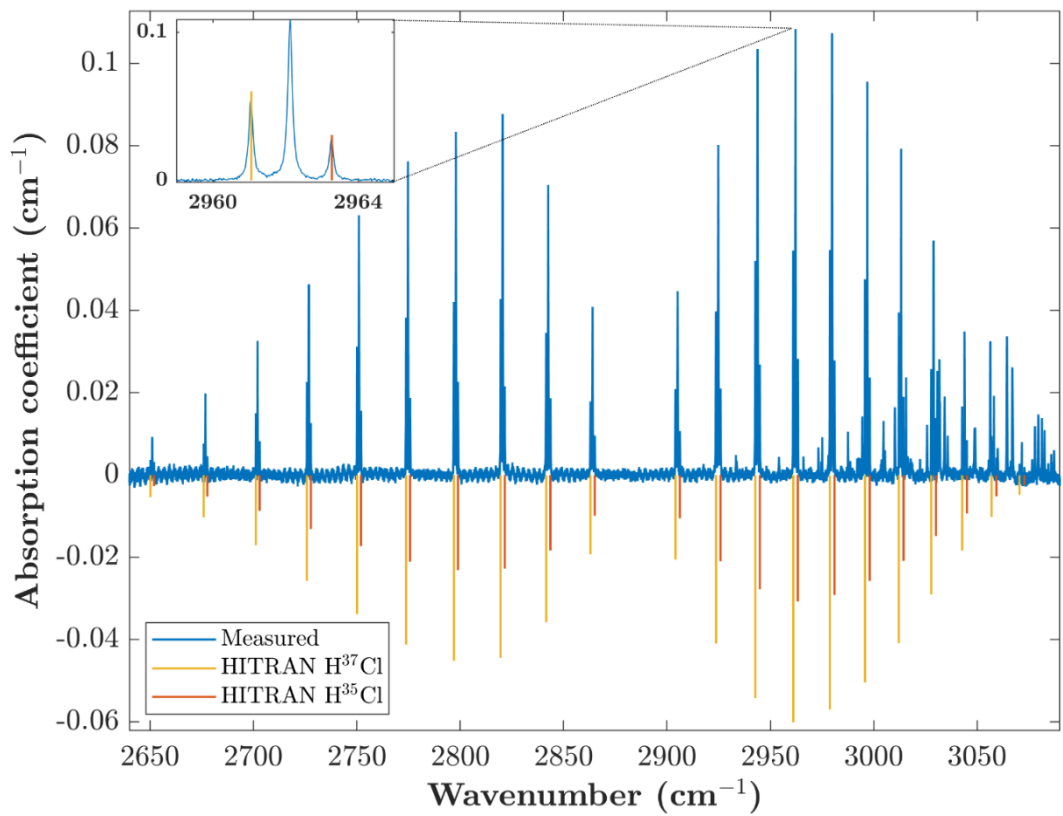

Fig. 1. The measured HCl absorption spectrum (blue). The middle peak of each of the peak triplets belongs to $H^{36}Cl$ as revealed by the included $H^{37}Cl$ (yellow) and $H^{35}Cl$ (orange) HITRAN lines (mirrored for clarity). The inset shows a zoomed view of the R(3) lines. In addition to the HCl peaks, there are interfering air absorption lines (especially in the R-branch), and an etalon effect caused by the sample cell windows. Note that the amounts of the stable isotopologues $H^{37}Cl$ and $H^{35}Cl$ in the sample do not follow their natural abundances. The $H^{35}Cl$ lines are thus the weakest in the spectrum.

The absorption spectrum in Fig. 1 shows a peak structure characteristic to the fundamental rovibrational band of HCl. Instead of the typical $H^{37}Cl$ and $H^{35}Cl$ peak doublets, we see triplets whose middle peak of each belongs to $H^{36}Cl$. We only consider lines P(10)–R(10) in this article, as lines further from the band center were masked by noise. The strongest $H^{36}Cl$ peak is line R(3) with 0.108 cm$^{-1}$ peak absorption (34 % in transmission) and the weakest is line R(10) with 0.008 cm$^{-1}$ peak absorption (92 % in transmission). In addition to the strong HCl lines, attention in Fig. 1 is drawn to the interfering air absorption lines and to the etalon effect. The free spectral range of the etalon is typically 2.3–2.8 cm$^{-1}$ (70–80 GHz), which matches well with the 1.25-mm thickness of the sample cell windows. The etalon and the interfering air absorption lines are included in the fit model as discussed below.

## 4. Fitting procedure

Due to the relatively high measurement pressure (1 atm), pressure broadening dominates most of the transition linewidths (see Supplementary Note 1). For this reason, we chose to model the absorption peaks as Lorentzian functions. All of the 21 rotational components (lines P(10)–R(10)) of the $H^{36}Cl$ fundamental band were fitted separately such that a typical fitting window was 4–9 cm$^{-1}$ around the $H^{36}Cl$ peak in question. To give an example, Fig. 2 shows the fit for line R(3). The fit model consists of three Lorentzian functions (one for each HCl isotopologue), a constant for a possible residual background offset, and a sine function to consider the etalon effect. In addition, extra Lorentzian functions were added for air absorption lines when necessary; in the case of the R(3) line, two small air peaks were fitted (Fig. 2).

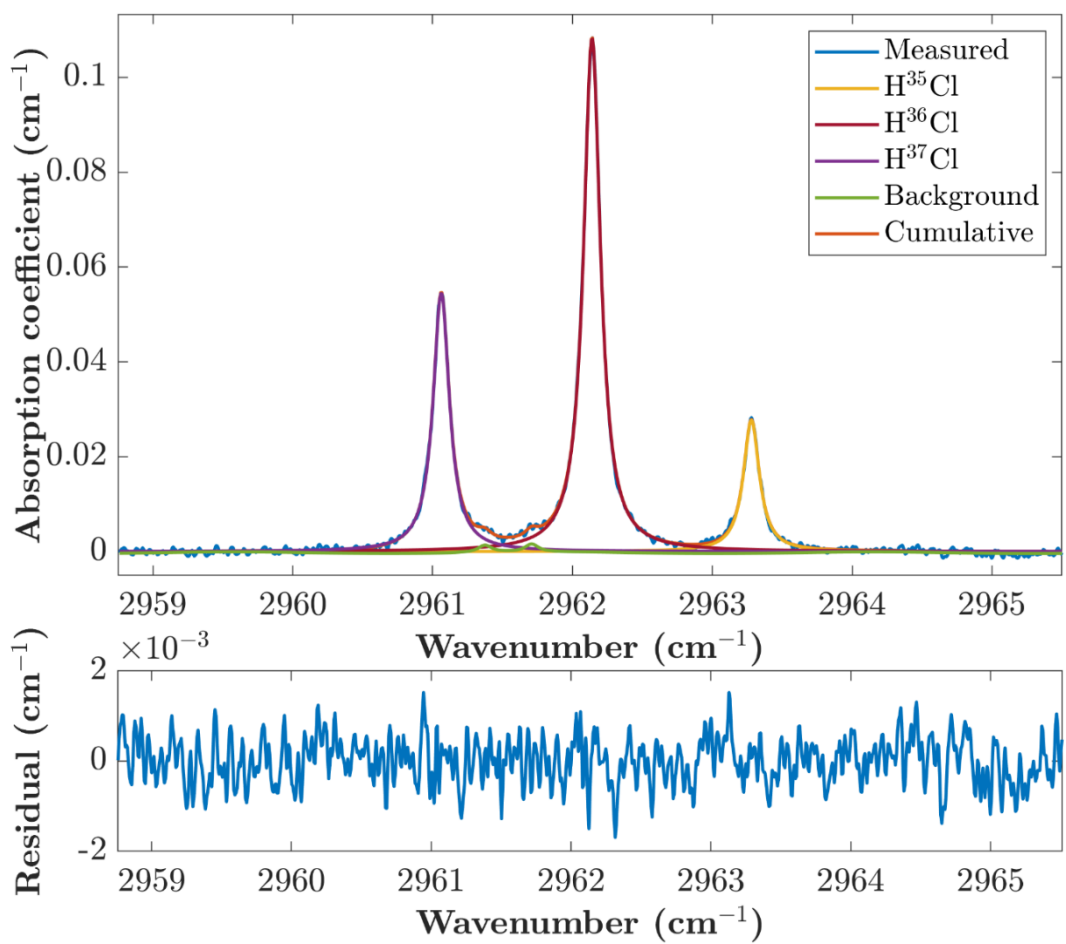

Fig. 2. Fit to the measured R(3) lines (upper panel) and the fit residual (lower panel). The three major peaks belong to $H^{37}Cl$ (leftmost), $H^{36}Cl$ (middle), and $H^{35}Cl$ (rightmost). The background fit consists of a constant, a sine function, and two Lorentzian functions for the two small air absorption peaks at approximately 2961.4 cm$^{-1}$ and 2961.7 cm$^{-1}$.

The lack of systematic features in the residual in Fig. 2 shows that the chosen fit model indeed describes the data well. The signal-to-noise ratio of the $H^{36}Cl$ peak is 212, defined as the ratio of the maximum peak absorption to the standard deviation of the fit residual. The peaks with the lowest SNR that were included in the analysis of the fundamental band are P(10) and R(10) (SNR≈15). Note that we also calibrated the wavenumber axis using the HITRAN zero pressure wavenumbers and the related 1-atm pressure shifts of the $H^{35}Cl$ and $H^{37}Cl$ P(10)–R(10) lines [20,54,55]. The analysis revealed an average offset of -0.0101 cm$^{-1}$ (-303 MHz) between the fitted $H^{35}Cl$ or $H^{37}Cl$ line center wavenumbers and the corresponding HITRAN values. Note that the average offset was calculated by weighting the data points by the inverses of the corresponding fit uncertainties. A similarly weighted standard deviation of the offsets is 0.00124 cm$^{-1}$ (37 MHz), which was combined with the fit uncertainties of the line center wavenumbers (see Results).

# 5. Results

Table 2 lists the fit results for the H$^{36}$Cl line center wavenumbers. The numbers in parentheses in Table 2 are one standard deviation uncertainties in least significant digits as obtained from the fit. The total uncertainty column contains the combined effect of the fit uncertainties and the calibration uncertainty at 68 % confidence level. Other uncertainty sources, such as the uncertainty of the baseline determination or the measurement pressure via the pressure shifts, are negligible. Overall, the total wavenumber uncertainties are below 0.0018 cm$^{-1}$ (60 MHz).

While determination of the line center wavenumbers was the main objective of this research, the measurements also provided information about the widths and relative intensities of the lines. These results are discussed separately in Supplementary information. Briefly, the relative intensities of the H$^{36}$Cl seem to follow those of the stable isotopologues H$^{35}$Cl and H$^{37}$Cl as expected based on previous research on the stable isotopologues. However, our results for the pressure-broadened linewidths are systematically larger than the HITRAN reference data for the stable isotopologues.

Table 2. Line center wavenumber $\tilde{\nu}$ results for the H$^{36}$Cl fundamental band. The numbers in parentheses are one standard deviation fit uncertainties in least significant digits and the values in the column denoted by $\sigma\tilde{\nu}$ represent the combined effect of the fit and calibration uncertainties at 68 % confidence level. The symbol $\tilde{\nu}_{\text{calc}}$ refers to the calculated line center wavenumbers as obtained from the rotational analysis discussed in the next section. Note that all the values in the table are given at 1 atm pressure and 296 K temperature.

| Line  | $\tilde{\nu}$ (cm$^{-1}$) | $\sigma\tilde{\nu}$ (10$^{-4}$ cm$^{-1}$) | $\tilde{\nu} - \tilde{\nu}_{\text{calc}}$ (10$^{-4}$ cm$^{-1}$) |
|-------|---------------------------|-------------------------------------------|-----------------------------------------------------------------|
| P(10) | 2651.0675(13)             | 18                                        | 9                                                               |
| P(9)  | 2676.8101(6)              | 14                                        | 0                                                               |
| P(8)  | 2702.0639(4)              | 13                                        | -6                                                              |
| P(7)  | 2726.8168(3)              | 13                                        | 3                                                               |
| P(6)  | 2751.0527(2)              | 13                                        | 10                                                              |
| P(5)  | 2774.7574(2)              | 13                                        | -1                                                              |
| P(4)  | 2797.9207(2)              | 13                                        | 0                                                               |
| P(3)  | 2820.5297(2)              | 13                                        | -5                                                              |
| P(2)  | 2842.5712(3)              | 13                                        | 7                                                               |
| P(1)  | 2864.0247(4)              | 13                                        | 1                                                               |
| R(0)  | 2905.1493(4)              | 13                                        | -1                                                              |
| R(1)  | 2924.7790(2)              | 13                                        | -2                                                              |
| R(2)  | 2943.7809(2)              | 13                                        | -4                                                              |
| R(3)  | 2962.1397(2)              | 13                                        | 6                                                               |
| R(4)  | 2979.8392(2)              | 13                                        | 1                                                               |
| R(5)  | 2996.8709(2)              | 13                                        | -1                                                              |
| R(6)  | 3013.2251(2)              | 13                                        | 7                                                               |
| R(7)  | 3028.8887(3)              | 13                                        | 4                                                               |
| R(8)  | 3043.8510(4)              | 13                                        | 3                                                               |
| R(9)  | 3058.1013(6)              | 14                                        | 2                                                               |
| R(10) | 3071.6279(13)             | 18                                        | -9                                                              |

## 6. Rotational analysis

We retrieved the molecular constants for the H$^{36}$Cl fundamental rovibrational band. The $v - 0$ band line center wavenumbers of a diatomic molecule can be approximately expressed as [56]

$$\tilde{v} = \tilde{v}_{v-0} + (\tilde{B}_v + \tilde{B}_0)m + (\tilde{B}_v - \tilde{B}_0 - \tilde{D}_v + \tilde{D}_0)m^2 + (\tilde{H}_v + \tilde{H}_0 - 2\tilde{D}_v - 2\tilde{D}_0)m^3$$
$$+ (3\tilde{H}_v - 3\tilde{H}_0 - \tilde{D}_v + \tilde{D}_0)m^4 + (3\tilde{H}_v + 3\tilde{H}_0)m^5 + (\tilde{H}_v - \tilde{H}_0)m^6, \quad (1)$$

where $\tilde{v}_{v-0}$ is the band center, $\tilde{B}$ is the rotational constant, and $\tilde{D}$ and $\tilde{H}$ are the different order centrifugal distortion constants. The integer $m = -J$ for the P-branch and $m = J + 1$ for the R-branch. The rotational quantum number $J$ refers to the lower state.

We could not determine the centrifugal distortion constants $\tilde{H}$ in a statistically significant manner; we thus assumed $\tilde{H} = \tilde{H}_v = \tilde{H}_0$ and constrained its value to $1.66742 \times 10^{-8}$ cm$^{-1}$. This value is the predicted $\tilde{H}_0$ of H$^{36}$Cl, which we calculated using $\tilde{H}_0$, $\tilde{H}_1$, and $\tilde{H}_2$ of H$^{35}$Cl given in the supplementary material of Ref. [20], the approximate Dunham expression [57,58]

$$\tilde{C}_{v,j} = \tilde{Y}_{0j} + \tilde{Y}_{1j}\left(v + \frac{1}{2}\right) + \tilde{Y}_{2j}\left(v + \frac{1}{2}\right)^2 \quad (2)$$

and the approximate isotope relation [15,67]

$$\frac{\tilde{Y}_{ij}^*}{\tilde{Y}_{ij}} = \rho^{i+2j}, \qquad \rho = \sqrt{\frac{\mu}{\mu^*}}, \quad (3)$$

where the asterisk denotes a different isotope, $\tilde{Y}_{ij}$ are Dunham coefficients [57], $\mu$ is the reduced mass, and $\tilde{C}_{v,0} - \tilde{C}_{0,0} = \tilde{v}_{v-0}$, $\tilde{C}_{v,1} = \tilde{B}_v$, $\tilde{C}_{v,2} = \tilde{D}_v$, and $\tilde{C}_{v,3} = \tilde{H}_v$.

Eqs. (2) and (3) and the molecular constants of H$^{35}$Cl given in the supplementary material of Ref. [20] were further used to predict in a similar manner the remaining molecular constants for H$^{36}$Cl. The results are listed in Table 4 together with the experimental molecular constants that we determined from the measured line center wavenumbers of lines P(10)–R(10) in least-squares fits using the combination relations [15,56]

$$R(J-1) - P(J+1) = \left(4\tilde{B}_0 - 6\tilde{D}_0 + \frac{27}{4}\tilde{H}\right)\left(J + \frac{1}{2}\right) - (8\tilde{D}_0 + 34\tilde{H})\left(J + \frac{1}{2}\right)^3 + 12\tilde{H}\left(J + \frac{1}{2}\right)^5 \quad (4)$$

$$R(J-1) + P(J) = 2\tilde{v}_{1-0} + 2(\tilde{B}_1 - \tilde{B}_0 - (\tilde{D}_1 - \tilde{D}_0))J^2 - 2(\tilde{D}_1 - \tilde{D}_0)J^4, \quad (5)$$

where $J = 1, \dots, 9$ for Eq. (4) and $J = 1, \dots, 10$ for Eq. (5).

First, the lower-state parameters $\tilde{B}_0$ and $\tilde{D}_0$ were determined in a least-squares fit using Eq. (4) and the assumption $\tilde{H} = 1.66742 \times 10^{-8}$ cm$^{-1}$. However, prior to fitting, we shifted the measured H$^{36}$Cl line center wavenumbers to those corresponding to zero pressure assuming the HITRAN H$^{35}$Cl pressure shifts [54,55]. After this, we determined the band center $\tilde{v}_{1-0}$ and upper-state parameters $\tilde{B}_1$ and $\tilde{D}_1$ using Eq. (5). The lower-state parameters were fixed during the least-squares fit to those obtained from the fit of Eq. (4). In addition, to consider the fit uncertainties of the lower-state parameters in the determination of the upper-state parameters, we repeated the fit a hundred times, each time drawing new $\tilde{B}_0$ and $\tilde{D}_0$ values from normal distributions whose expectation values and standard deviations were chosen to be the original $\tilde{B}_0$ and $\tilde{D}_0$ values and their respective fit uncertainties obtained from the fit of Eq. (4). Finally, we calculated the standard deviations of the hundred new $\tilde{B}_0$ and $\tilde{D}_0$ values and combined them with the fit uncertainties obtained from the original fit of Eq. (5) to estimate the total uncertainties of the upper-state parameters.

Table 3. The experimental and predicted fundamental band molecular constants for H$^{36}$Cl obtained as explained in the text. The values refer to zero pressure.

| Parameter | Experimental (cm$^{-1}$) | Predicted (cm$^{-1}$) |
| --- | --- | --- |
| $\tilde{v}_{1-0}$ | 2884.8936(3) | 2884.8933 |
| $\tilde{B}_0$ | 10.43210(4) | 10.43213 |
| $\tilde{B}_1$ | 10.12843(4) | 10.12847 |
| $\tilde{D}_0/10^{-4}$ | 5.275(3) | 5.273 |
| $\tilde{D}_1/10^{-4}$ | 5.207(3) | 5.206 |
| $\tilde{H}_0/10^{-8}$ |  | 1.66742 |

The experimental and predicted molecular constants in Table 4 are in excellent agreement. This is also true for the observed minus calculated line center wavenumber values in column 4 of Table 2 (denoted by $\tilde{v} - \tilde{v}_{\text{calc}}$). These values were calculated using Eq. (1), the experimental constants in Table 4, and the assumption $\tilde{H} = \tilde{H}_v = \tilde{H}_0$; the pressure shifts were reapplied before the comparison. The largest difference is 0.0010 cm$^{-1}$ (30 MHz) and all the values are within the estimated total uncertainties.

Although the experimental and calculated wavenumbers are in excellent agreement, it is well known that the molecular parameters obtained with this traditional approach may perform poorly in predicting accurate line center wavenumbers for transitions with high quantum number $J$ values [20]. Indeed, we repeated the above-explained procedure to determine the molecular constants also for H$^{35}$Cl and H$^{37}$Cl. Even though the result parameters were within two standard deviations from those obtained with more refined methods [20] or from highly accurate experiments consisting of a considerably larger dataset [59], our model for H$^{35}$Cl (H$^{37}$Cl) predicts up to 0.050 cm$^{-1}$ or 1.5 GHz larger (0.020 cm$^{-1}$ or 0.6 GHz lower) wavenumber values for lines P(15) and R(15) when compared to the corresponding HITRAN line center wavenumbers. We can then expect similar inaccuracy for the H$^{36}$Cl model.

## 7. The first overtone band

We also measured the first overtone 2 − 0 rovibrational band of H$^{36}$Cl. Due to the limited amount of Na$^{36}$Cl solution available for us at the time, we conducted the measurements with the same sample that we used in the MIR measurements. For these near infrared (NIR) measurements, we co-added 100 interferograms and used a Si-on-CaF$_2$ beam splitter, Tungsten light source, and a liquid-nitrogen-cooled InSb detector. The measurement settings, baseline determination, and fitting procedure were similar to what we used in the MIR measurements. The relevant measurement settings are listed in Table 5.

Table 4. Relevant FTIR instrument settings for the NIR measurements.

| Light source | Tungsten |
| --- | --- |
| Beam splitter | Si-on-CaF$_2$ |
| Detector | InSb |
| Aperture diameter | 1 mm |
| Apodization function | Norton-Beer medium |
| HWHM of ILS function | 0.0095 cm$^{-1}$ (285 MHz) |
| Number of co-added interferograms | 100 |
| Type of interferograms | Double-sided |
| Interferogram processing method | Mertz |
| Zero-filling factor | 2 |
| Optical band-pass filtering | 6667-5000 cm$^{-1}$ |
| Digital band-pass filtering | 25–35 kHz (4937-6912 cm$^{-1}$) |
| HeNe reference laser down-converted frequency | 80 kHz |

Despite the longer averaging time, the SNR in the NIR measurements is worse than in the MIR measurements. This was expected, since the HCl overtone transitions are typically 30–50 times weaker than the corresponding fundamental ones. The P-branch is also considerably interfered by air absorption. For these reasons, we could only reliably analyze lines P(1)–R(7), and only the $H^{37}Cl$ lines R(1)–R(7) were used in the wavenumber axis calibration. The calibration revealed an average offset of -0.01479 $cm^{-1}$ (-443 MHz) compared to the corresponding HITRAN values [20,54]. However, because the HITRAN pressure-shift data is extrapolated from the MIR measurements by Pine and Looney [55], we chose to use the NIR pressure-shift data by Asfin et al. [23] to calculate the $H^{37}Cl$ line center wavenumbers at 1 atm. The standard deviation of the offsets is 0.00364 $cm^{-1}$ (109 MHz) and was again interpreted as the calibration uncertainty. Note that in the calculation of the average and the standard deviation of the offsets, the data points were again weighted using the inverses of the fit uncertainties. As an example, Fig. 5 shows the overtone line R(3) fit. This is the strongest $H^{36}Cl$ line whose peak absorption is 0.0027 $cm^{-1}$ (97.3% in transmission) with an SNR of 20; the lowest SNR is 7 and for line P(1) with 0.0010 $cm^{-1}$ peak absorption (99.0% in transmission).

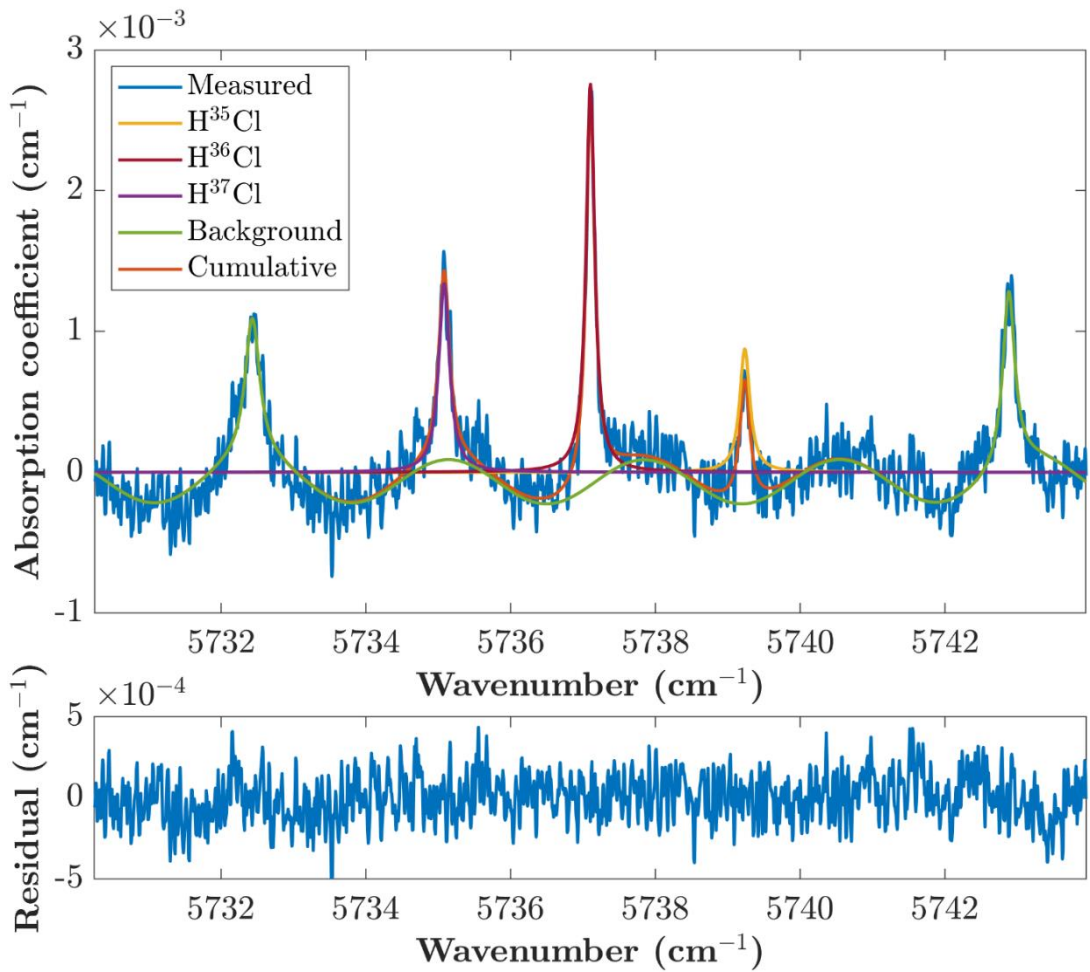

Fig. 3. Fit to the measured first overtone R(3) lines (upper panel) and the fit residual (lower panel). The leftmost and rightmost peaks are fitted air absorption lines and the remaining three fitted peaks belong to $H^{37}Cl$, $H^{36}Cl$, and $H^{35}Cl$ in the order of increasing wavenumber. In addition to the two air absorption lines, the background function consists of a constant and a sine.

Table 6 lists the fit parameter results for the H$^{36}$Cl first overtone band. Due to the generally modest SNR, we could not determine the linewidths and relative intensities reliably, for which reason we report only the line center wavenumbers. The numbers in parentheses are the one standard deviation fit uncertainties in least significant digits. The total uncertainty column values represent the combined fit and calibration uncertainties at 68 % confidence level. They are less than 0.007 cm$^{-1}$ (0.2 GHz). The observed minus calculated column values in Table 6 are again in good agreement with the experiment and deviate less than 0.005 cm$^{-1}$ (150 MHz) from the experimental values, not including the low SNR lines P(1) and R(0), for which large deviations are apparent. The observed minus calculated values have been obtained from a similar rotational analysis to what was discussed for the MIR measurements. However, due to the small number of P-branch lines, we used Eq. (1) directly in the least-squares fit, made the $\widetilde{H}_v = \widetilde{H}_0 = 1.66742 \times 10^{-8}$ cm$^{-1}$ assumption, and fixed the remaining lower-state parameters to the experimental values in Table 4.

Table 5. Line parameter results for the H$^{36}$Cl first overtone band. The line center wavenumber is denoted by $\tilde{\nu}$. The numbers in parentheses are one standard deviation fit uncertainties in least significant digits and the values in the column denoted by $\sigma\tilde{\nu}$ represent the combined effect of the fit and calibration uncertainties at 68 % confidence level. The symbol $\tilde{\nu}_{\text{calc}}$ refers to the calculated line center wavenumbers as obtained from the rotational analysis discussed in the text. Note that all the wavenumber values in the table are given at 1 atm pressure.

| Line | $\tilde{\nu}$ (cm$^{-1}$) | $\sigma\tilde{\nu}$ (10$^{-3}$ cm$^{-1}$) | $\tilde{\nu} - \tilde{\nu}_{\text{calc}}$ (10$^{-3}$ cm$^{-1}$) |
|---|---|---|---|
| P(1) | 5645.043(5) | 7 | 21 |
| R(0) | 5685.525(5) | 6 | -21 |
| R(1) | 5703.965(3) | 5 | -5 |
| R(2) | 5721.168(2) | 5 | 3 |
| R(3) | 5737.113(2) | 4 | -1 |
| R(4) | 5751.802(2) | 4 | -1 |
| R(5) | 5765.227(2) | 4 | 2 |
| R(6) | 5777.370(3) | 5 | 3 |
| R(7) | 5788.217(4) | 5 | -2 |

Table 7 lists the experimental and predicted molecular constants obtained for the first overtone band in a similar manner as discussed for the fundamental band. The accuracy of the experimental parameters is not sufficient to ensure accurate line center wavenumber predictions for high $J$ values. We advise using the predicted molecular constants instead. For example, similarly predicted molecular constants for H$^{37}$Cl predict up to 0.027 cm$^{-1}$ (0.8 GHz) larger line center wavenumbers for lines P(15) and R(15) compared to HITRAN. We note that the predicted molecular constants for H$^{37}$Cl agree within at least five significant decimals compared to the literature values [20] not including the band center, which agrees within 0.0011 cm$^{-1}$ (33 MHz).

Table 6. Experimental and predicted first overtone band molecular constants for H$^{36}$Cl obtained as explained in the text. The values refer to zero pressure.

| Parameter | Experimental (cm$^{-1}$) | Predicted (cm$^{-1}$) |
|---|---|---|
| $\tilde{\nu}_{2-0}$ | 5665.895(10) | 5665.898 |
| $\widetilde{B}_2$ | 9.8274(8) | 9.8272 |
| $\widetilde{D}_2/10^{-4}$ | 5.18(11) | 5.15 |

# 8. Conclusions and outlook

In this article, we reported the first absorption spectrum of H$^{36}$Cl. The measurements of the fundamental and the first overtone rovibrational bands were performed using a commercial FTIR instrument with a thermal light source and using a sample prepared from the reaction between sulfuric acid and $^{36}$Cl-enriched NaCl salt. The simple gas cell system helped to maximize the number density of the species of interest inside the cell, which provided us with strong absorption lines in the MIR and allowed us to determine the fundamental band P(10)–R(10) line center wavenumbers (referenced to the HITRAN values of the stable isotopologues H$^{35}$Cl and H$^{37}$C), with total uncertainties of less than 0.0018 cm$^{-1}$ (60 MHz) at 68 % confidence level. Due to the order of magnitude weaker molecular transitions and consequently modest SNR in the NIR, we could determine the first overtone band line center wavenumbers of only lines P(1)–R(7) and with total uncertainties of less than 0.007 cm$^{-1}$ (0.2 GHz) at 68 % confidence level.

The simple gas cell system came with the downside of not knowing the exact pressure and temperature inside the sample cell. However, the initiation of the reaction at atmospheric pressure and room temperature made it possible to estimate the measurement conditions with reasonable accuracy. In the Supplementary information, we investigated the linewidths and relative intensities of the H$^{36}$Cl lines and compared them to the corresponding literature values of the stable isotopologues H$^{35}$Cl and H$^{37}$Cl. The relative intensities showed similar behavior as the reference HITRAN data, implying that the line intensities of the different isotopologues are similar as expected based on previous research on the stable isotopologues. However, as discussed in the Supplementary information, the linewidths for all three isotopologues were systematically larger than the HITRAN reference data for H$^{35}$Cl and H$^{37}$Cl.

An advantage of having atmospheric pressure inside the sample cell during measurements was to minimize the distorting effect of the ILS function and to allow us to use a simple Lorentzian fit model. However, precise control of the measurement pressure and measurements in lower pressures to reduce the linewidths would enable much more accurate line center wavenumber determinations. To increase measurement sensitivity, an external coherent light source in the FTIR instrument instead of the thermal source could be useful, especially if combined with photoacoustic detection [47]. For accurate linewidth determinations, eliminating the ILS function altogether would be helpful. This could be done by using optical frequency combs [60] or by avoiding FTIR altogether (i.e., by using tunable laser absorption spectroscopy instead) [55,61,62]. In particular, the sensitivity improvement by laser absorption spectroscopy would allow more accurate characterization of the first overtone band. In addition, determination of the absolute line intensities would require a more elaborate sample preparation and gas-control system to ensure accurate external knowledge of the sample concentration [22]. Finally, as future applications in monitoring H$^{36}$Cl will expectedly require very high sensitivities to detect trace amounts of this species, it would be valuable to demonstrate low-concentration measurements using some highly sensitive laser spectroscopy method, such as cavity ring-down spectroscopy [34,39] or photoacoustic spectroscopy [37,46,63]. These pursuits are assisted by our results, as we now have reasonably accurate knowledge on the line center wavenumbers of H$^{36}$Cl, and the simple models from the rotational analyses can be used to predict line positions of yet unobserved transitions.

## Acknowledgements

We wish to thank Dr. Markus Metsälä and Prof. Lauri Halonen for their support and constructive feedback on the project, and Dr. Guillaume Genoud for fruitful discussions on the topic and for the work to secure funding for the project.

## Funding

This work obtained funding from the Euratom research and training programme 2014-2018 [grant agreement number 755371, the CHANCE project]; the Academy of Finland [project number 326444]; and the Flagship of Photonics Research and Innovation of the Academy of Finland (PREIN). S. Larnimaa acknowledges financial support from the CHEMS doctoral program of the University of Helsinki.

## Author contributions

Santeri Larnimaa: Investigation, Formal analysis, Writing – Original Draft, Writing – Review & Editing, Visualization. Markku Vainio: Conceptualization, Writing – Review & Editing, Supervision, Project Administration. Ville Ulvila: Conceptualization, Investigation, Writing – Review & Editing, Supervision, Project Administration.

## References

1. W. R. Simpson, S. S. Brown, A. Saiz-Lopez, J. A. Thornton and R. von Glasow, "Tropospheric Halogen Chemistry: Sources, Cycling, and Impacts", *Chem. Rev.*, 2015, **115**, 10, pp. 4035-4062, https://doi.org/10.1021/cr5006638.

2. M. A. Zondlo, P. K. Hudson, A. J. Prenni and M. A. Tolbert, "Chemistry and Microphysics of Polar Stratospheric Clouds and Cirrus Clouds", *Annu. Rev. Phys. Chem.*, 2000, **51**, 1, pp. 473-499, https://doi.org/10.1146/annurev.physchem.51.1.473.

3. F. Lique and A. Faure, "Collisional excitation and dissociation of HCl by H", *Mon. Not. R. Astron. Soc.*, 2017, **472**, 1, pp. 738-743, https://doi.org/10.1093/mnras/stx2025.

4. M. Lanza, Y. Kalugina, L. Wiesenfeld, A. Faure and F. Lique, "New insights on the HCl abundance in the interstellar medium", *Mon. Not. R. Astron. Soc.*, 2014, **443**, 4, pp. 3351-3358, https://doi.org/10.1093/mnras/stu1371.

5. R. Peng, H. Yoshida, R. A. Chamberlin, T. G. Phillips, D. C. Lis and M. Gerin, "A COMPREHENSIVE SURVEY OF HYDROGEN CHLORIDE IN THE GALAXY", *Astrophys. J.*, 2010, **723**, 1, pp. 218-228, https://doi.org/10.1088/0004-637x/723/1/218.

6. H. Panu, R. Timo, F. Thomas, J. Makkonen, S. Alyshev, A. Kharakhordin and S. Firstov, "Real-time HCl gas detection at parts-per-billion level concentrations utilising a diode laser and a bismuth-doped fibre amplifier", *Meas. Sci. Technol.*, 2021, **32**, 055206, https://doi.org/10.1088/1361-6501/abd651.

7. V. V. Zuev, N. E. Zueva, E. S. Savelieva and V. V. Gerasimov, "The Antarctic ozone depletion caused by Erebus volcano gas emissions", *Atmos. Environ.*, 2015, **122**, pp. 393-399, https://doi.org/10.1016/j.atmosenv.2015.10.005.

8. G. Tamburello, A. Caselli, F. Tassi, O. Vaselli, S. Calabrese, D. Rouwet, B. Capaccioni, R. Di Napoli, C. Cardellini, G. Chiodini, M. Bitetto, L. Brusca, S. Bellomo and A. Aiuppa, "Intense magmatic degassing through the lake of Copahue volcano, 2013–2014", *J. Geophys. Res. Solid Earth*, 2015, **120**, 9, pp. 6071-6084, https://doi.org/10.1002/2015JB012160.

9. C. Voigt, P. Jessberger, T. Jurkat, S. Kaufmann, R. Baumann, H. Schlager, N. Bobrowski, G. Giuffrida and G. Salerno, "Evolution of $CO_2$, $SO_2$, HCl, and $HNO_3$ in the volcanic plumes from Etna", *Geophys. Res. Lett.*, 2014, **41**, 6, pp. 2196-2203, https://doi.org/10.1002/2013GL058974.

10. D. M. Pyle and T. A. Mather, "Halogens in igneous processes and their fluxes to the atmosphere and oceans from volcanic activity: A review", *Chem. Geol.*, 2009, **263**, 1, pp. 110-121, https://doi.org/10.1016/j.chemgeo.2008.11.013.

11. O. Korablev, K. S. Olsen, A. Trokhimovskiy, F. Lefèvre, F. Montmessin, A. A. Fedorova, M. J. Toplis, J. Alday, D. A. Belyaev, A. Patrakeev, N. I. Ignatiev, A. V. Shakun, A. V. Grigoriev, L. Baggio, I. Abdenour, G. Lacombe, Y. S. Ivanov, S. Aoki, I. R. Thomas, F. Daerden, B. Ristic, J. T. Erwin, M. Patel, G. Bellucci, J. Lopez-Moreno and A. C. Vandaele, "Transient HCl in the atmosphere of Mars", *Sci. Adv.*, 2021, **7**, 7, eabe4386, https://doi.org/10.1126/sciadv.abe4386.

12. V. A. Krasnopolsky, D. A. Belyaev, I. E. Gordon, G. Li and L. S. Rothman, "Observations of D/H ratios in $H_2O$, HCl, and HF on Venus and new DCl and DF line strengths", *Icarus*, 2013, **224**, 1, pp. 57-65, https://doi.org/10.1016/j.icarus.2013.02.010.

13. B. J. Sandor and R. T. Clancy, "Observations of HCl altitude dependence and temporal variation in the 70–100 km mesosphere of Venus", *Icarus*, 2012, **220**, 2, pp. 618-626, https://doi.org/10.1016/j.icarus.2012.05.016.

14. D. H. Rank, B. S. Rao and T. A. Wiggins, "Molecular constants of $HCl^{35}$", *J. Mol. Spectrosc.*, 1965, **17**, 1, pp. 122-130, https://doi.org/10.1016/0022-2852(65)90114-1.

15. D. H. Rank, D. P. Eastman, B. S. Rao and T. A. Wiggins, "Rotational and Vibrational Constants of the $HCl^{35}$ and $DCl^{35}$ Molecules*", *J. Opt. Soc. Am.*, 1962, **52**, 1, pp. 1-7, https://doi.org/10.1364/JOSA.52.000001.

16. D. H. Rank, W. B. Birtley, D. P. Eastman, B. S. Rao and T. A. Wiggins, "Precise Measurements of Some Infrared Bands of Hydrogen Chloride*", *J. Opt. Soc. Am.*, 1960, **50**, 12, pp. 1275-1279, https://doi.org/10.1364/JOSA.50.001275.

17. E. D. Palik, "History of far-infrared research. I. The Rubens era", *J. Opt. Soc. Am.*, 1977, **67**, 7, pp. 857-865, https://doi.org/10.1364/JOSA.67.000857.

18. N. Ginsburg, "History of far-infrared research. II. The grating era, 1925–1960", *J. Opt. Soc. Am.*, 1977, **67**, 7, pp. 865-871, https://doi.org/10.1364/JOSA.67.000865.

19. E. S. Imes, "Measurements on the Near Infra-Red Absorption of Some Diatomic Gases", *Astrophys. J.*, 1919, **50**, pp. 251-276, https://doi.org/10.1086/142504.

20. J. A. Coxon and P. G. Hajigeorgiou, "Improved direct potential fit analyses for the ground electronic states of the hydrogen halides: HF/DF/TF, HCl/DCl/TCl, HBr/DBr/TBr and HI/DI/TI", *J. Quant. Spectrosc. Radiat. Transfer*, 2015, **151**, pp. 133-154, https://doi.org/10.1016/j.jqsrt.2014.08.028.

21. G. Li, I. E. Gordon, P. F. Bernath and L. S. Rothman, "Direct fit of experimental ro-vibrational intensities to the dipole moment function: Application to HCl", *J. Quant. Spectrosc. Radiat. Transfer*, 2011, **112**, 10, pp. 1543-1550, https://doi.org/10.1016/j.jqsrt.2011.03.014.

22. G. Li, A. Serdyukov, M. Gisi, O. Werhahn and V. Ebert, "FTIR-based measurements of self-broadening and self-shift coefficients as well as line strength in the first overtone band of HCl at 1.76 μM", *J. Quant. Spectrosc. Radiat. Transfer*, 2015, **165**, pp. 76-87, https://doi.org/10.1016/j.jqsrt.2015.06.021.

23. R. E. Asfin, A. V. Domanskaya and C. Maul, "Broadening and shifting coefficients of rotation–vibrational lines in the fundamental and first overtone bands of HCl and HBr induced by oxygen and air", *J. Quant. Spectrosc. Radiat. Transfer*, 2013, **130**, pp. 296-303, https://doi.org/10.1016/j.jqsrt.2013.07.014.

24. K. Iwakuni, H. Sera, M. Abe and H. Sasada, "Hyperfine-resolved transition frequency list of fundamental vibration bands of H$^{35}$Cl and H$^{37}$Cl", *J. Mol. Spectrosc.*, 2014, **306**, pp. 19-25, https://doi.org/10.1016/j.jms.2014.09.013.

25. D. Huggle, A. Blinov, C. Stan-Sion, G. Korschinek, C. Scheffel, S. Massonet, L. Zerle, J. Beer, Y. Parrat, H. Gaeggeler, W. Hajdas and E. Nolte, "Production of cosmogenic $^{36}$Cl on atmospheric argon", *Planet. Space Sci.*, 1996, **44**, 2, pp. 147-151, https://doi.org/10.1016/0032-0633(95)00085-2.

26. Y. Tosaki, N. Morikawa, K. Kazahaya, H. Tsukamoto, Y. S. Togo, T. Sato, H. A. Takahashi, M. Takahashi and A. Inamura, "Deep incursion of seawater into the Hiroshima Granites during the Holocene transgression: Evidence from $^{36}$Cl age of saline groundwater in the Hiroshima area, Japan", *Geochem. J.*, 2017, **51**, 3, pp. 263-275, https://doi.org/10.2343/geochemj.2.0467.

27. V. Lavastre, C. Le Gal La Salle, J. Michelot, S. Giannesini, L. Benedetti, J. Lancelot, B. Lavielle, M. Massault, B. Thomas, E. Gilabert, D. Bourlès, N. Clauer and P. Agrinier, "Establishing constraints on groundwater ages with $^{36}$Cl, $^{14}$C, $^{3}$H, and noble gases: A case study in the eastern Paris basin, France", *Appl. Geochem.*, 2010, **25**, 1, pp. 123-142, https://doi.org/10.1016/j.apgeochem.2009.10.006.

28. Y. Tosaki, N. Tase, G. Massmann, Y. Nagashima, R. Seki, T. Takahashi, K. Sasa, K. Sueki, T. Matsuhiro, T. Miura, K. Bessho, H. Matsumura and M. He, "Application of $^{36}$Cl as a dating tool for modern groundwater", *Nucl. Instrum. Methods Phys. Res., Sect. B*, 2007, **259**, 1, pp. 479-485, https://doi.org/10.1016/j.nimb.2007.02.096.

29. A. J. Love, A. L. Herczeg, L. Sampson, R. G. Cresswell and L. K. Fifield, "Sources of chloride and implications for $^{36}$Cl dating of old groundwater, Southwestern Great Artesian Basin, Australia", *Water Resour. Res.*, 2000, **36**, 6, pp. 1561-1574, https://doi.org/10.1029/2000WR900019.

30. S. Le Dizès and M. A. Gonze, "Behavior of $^{36}$Cl in agricultural soil-plant systems: A review of transfer processes and modelling approaches", *J. Environ. Radioact.*, 2019, **196**, pp. 82-90, https://doi.org/10.1016/j.jenvrad.2018.10.011.

31. T. L. White, D. DiPrete, C. DiPrete and G. Dobos, "Analysis of $^{36}$Cl in Savannah River Site radioactive waste", *J. Radioanal. Nucl.*, 2013, **296**, 2, pp. 835-839, https://doi.org/10.1007/s10967-012-2071-9.

32. S. C. Sheppard, L. H. Johnson, B. W. Goodwin, J. C. Tait, D. M. Wuschke and C. C. Davison, "Chlorine-36 in nuclear waste disposal—1. Assessment results for used fuel with comparison to $^{129}$I and $^{14}$C", *Waste Manage.*, 1996, **16**, 7, pp. 607-614, https://doi.org/10.1016/S0956-053X(97)00001-9.

33. T. M. Beasley, D. Elmore, P. W. Kubik and P. Sharma, "Chlorine-36 Releases from the Savanniah River Site Nuclear Fuel Reprocessing Facilities", *Groundwater*, 1992, **30**, 4, pp. 539-548, https://doi.org/10.1111/j.1745-6584.1992.tb01530.x.

34. I. Galli, S. Bartalini, R. Ballerini, M. Barucci, P. Cancio, M. De Pas, G. Giusfredi, D. Mazzotti, N. Akikusa and P. De Natale, "Spectroscopic detection of radiocarbon dioxide at parts-per-quadrillion sensitivity", *Optica*, 2016, **3**, 4, pp. 385-388, https://doi.org/10.1364/OPTICA.3.000385.

35. W. Kutschera, "Applications of accelerator mass spectrometry", *Int. J. Mass Spectrom.*, 2013, **349-350**, pp. 203-218, https://doi.org/10.1016/j.ijms.2013.05.023.

36. X. Hou, "Liquid scintillation counting for determination of radionuclides in environmental and nuclear application", *J. Radioanal. Nucl.*, 2018, **318**, 3, pp. 1597-1628, https://doi.org/10.1007/s10967-018-6258-6.

37. M. Fatima, T. Hausmaninger, T. Tomberg, J. Karhu, M. Vainio, T. Hieta and G. Genoud, "Radiocarbon dioxide detection using cantilever-enhanced photoacoustic spectroscopy", *Opt. Lett.*, 2021, **46**, 9, pp. 2083-2086, https://doi.org/10.1364/OL.420199.

38. R. Terabayashi, K. Saito, V. Sonnenschein, Y. Okuyama, T. Iguchi, M. Yamanaka, N. Nishizawa, K. Yoshida, S. Ninomiya and H. Tomita, "Mid-infrared cavity ring-down spectroscopy using DFB quantum cascade laser with optical feedback for radiocarbon detection", *Jpn. J. Appl. Phys.*, 2020, **59**, 9, 092007, https://doi.org/10.35848/1347-4065/abb20e.

39. G. Genoud, J. Lehmuskoski, S. Bell, V. Palonen, M. Oinonen, M. Koskinen-Soivi and M. Reinikainen, "Laser Spectroscopy for Monitoring of Radiocarbon in Atmospheric Samples", *Anal. Chem.*, 2019, **91**, 19, pp. 12315-12320, https://doi.org/10.1021/acs.analchem.9b02496.

40. V. Sonnenschein, R. Terabayashi, H. Tomita, S. Kato, N. Hayashi, S. Takeda, L. Jin, M. Yamanaka, N. Nishizawa, A. Sato, K. Yoshida and T. Iguchi, "A cavity ring-down spectrometer for study of biomedical radiocarbon-labeled samples", *J. Appl. Phys.*, 2018, **124**, 3, 033101, https://doi.org/10.1063/1.5041015.

41. A. J. Fleisher, D. A. Long, Q. Liu, L. Gameson and J. T. Hodges, "Optical Measurement of Radiocarbon below Unity Fraction Modern by Linear Absorption Spectroscopy", *J. Phys. Chem. Lett.*, 2017, **8**, 18, pp. 4550-4556, https://doi.org/10.1021/acs.jpclett.7b02105.

42. A. D. McCartt, T. J. Ognibene, G. Bench and K. W. Turteltaub, "Quantifying Carbon-14 for Biology Using Cavity Ring-Down Spectroscopy", *Anal. Chem.*, 2016, **88**, 17, pp. 8714-8719, https://doi.org/10.1021/acs.analchem.6b02054.

43. G. Genoud, M. Vainio, H. Phillips, J. Dean and M. Merimaa, "Radiocarbon dioxide detection based on cavity ring-down spectroscopy and a quantum cascade laser", *Opt. Lett.*, 2015, **40**, 7, pp. 1342-1345, https://doi.org/10.1364/OL.40.001342.

44. I. Galli, S. Bartalini, S. Borri, P. Cancio, D. Mazzotti, P. De Natale and G. Giusfredi, "Molecular Gas Sensing Below Parts Per Trillion: Radiocarbon-Dioxide Optical Detection", *Phys. Rev. Lett.*, 2011, **107**, 27, 270802, https://doi.org/10.1103/PhysRevLett.107.270802.

45. I. Galli, P. C. Pastor, G. Di Lonardo, L. Fusina, G. Giusfredi, D. Mazzotti, F. Tamassia and P. De Natale, "The $v_3$ band of $^{14}C^{16}O_2$ molecule measured by optical-frequency-comb-assisted cavity ring-down spectroscopy", *Mol. Phys.*, 2011, **109**, 17-18, pp. 2267-2272, https://doi.org/10.1080/00268976.2011.614284.

46. S. Larnimaa, L. Halonen, J. Karhu, T. Tomberg, M. Metsälä, G. Genoud, T. Hieta, S. Bell and M. Vainio, "High-resolution analysis of the $v_3$ band of radiocarbon methane $^{14}CH_4$", *Chem. Phys. Lett.*, 2020, **750**, 137488, https://doi.org/10.1016/j.cplett.2020.137488.

47. J. Karhu, T. Tomberg, F. Senna Vieira, G. Genoud, V. Hänninen, M. Vainio, M. Metsälä, T. Hieta, S. Bell and L. Halonen, "Broadband photoacoustic spectroscopy of $^{14}CH_4$ with a high-power mid-infrared optical frequency comb", *Opt. Lett.*, 2019, **44**, 5, pp. 1142-1145, https://doi.org/10.1364/OL.44.001142.

48. M. Yim and F. Caron, "Life cycle and management of carbon-14 from nuclear power generation", *Prog. Nucl. Energy*, 2006, **48**, 1, pp. 2-36, https://doi.org/10.1016/j.pnucene.2005.04.002.

49. J. I. Gmitro and T. Vermeulen, "Vapor-liquid equilibria for aqueous sulfuric acid", *AIChE J.*, 1964, **10**, 5, pp. 740-746, https://doi.org/10.1002/aic.690100531.

50. P. R. Griffiths and J. A. de Haseth, *Fourier Transform Infrared Spectrometry, Chemical Analysis: A Series of Monographs on Analytical Chemistry and Its Applications*, Vol. 83, eds. P. J. Elving, J. D. Winefordner and I. M. Kolthoff, John Wiley & Sons, 1986.

51. R. H. Norton and R. Beer, "New apodizing functions for Fourier spectrometry", *J. Opt. Soc. Am.*, 1976, **66**, 3, pp. 259-264, https://doi.org/10.1364/JOSA.66.000259.

52. A. Savitzky and M. J. E. Golay, "Smoothing and Differentiation of Data by Simplified Least Squares Procedures", *Anal. Chem.*, 1964, **36**, 8, pp. 1627-1639, https://doi.org/10.1021/ac60214a047.

53. P. F. Bernath, *Spectra of Atoms and Molecules*, Oxford University Press, New York, 1995.

54. I. E. Gordon, L. S. Rothman, C. Hill, R. V. Kochanov, Y. Tan, P. F. Bernath, M. Birk, V. Boudon, A. Campargue, K. V. Chance, B. J. Drouin, J.-M. Flaud, R. R. Gamache, J. T. Hodges, D. Jacquemart, V. I. Perevalov, A. Perrin, K. P. Shine, M.-A. H. Smith, J. Tennyson, G. C. Toon, H. Tran, V. G. Tyuterev, A. Barbe, A. G. Császár, V. M. Devi, T. Furtenbacher, J. J. Harrison, J.-M. Hartmann, A. Jolly, T. J. Johnson, T. Karman, I. Kleiner, A. A. Kyuberis, J. Loos, O. M. Lyulin, S. T. Massie, S. N. Mikhailenko, N. Moazzen-Ahmadi, H. S. P. Müller, O. V. Naumenko, A. V. Nikitin, O. L. Polyansky, M. Rey, M. Rotger, S. W. Sharpe, K. Sung, E. Starikova, S. A. Tashkun, J. V. Auwera, G. Wagner, J. Wilzewski, P. Wcisło, S. Yu and E. J. Zak, "The HITRAN2016 molecular spectroscopic database", *J. Quant. Spectrosc. Radiat. Transfer*, 2017, **203**, pp. 3-69, https://doi.org/10.1016/j.jqsrt.2017.06.038.

55. A. S. Pine and J. P. Looney, "$N_2$ and air broadening in the fundamental bands of HF and HCl", *J. Mol. Spectrosc.*, 1987, **122**, 1, pp. 41-55, https://doi.org/10.1016/0022-2852(87)90217-7.

56. G. Herzberg, *Molecular Spectra and Molecular Structure I. Spectra of Diatomic Molecules*, 2nd ed., D. Van Nostrand Company, Inc., Princeton, 1950.

57. J. L. Dunham, "The Energy Levels of a Rotating Vibrator", *Phys. Rev.*, 1932, **41**, 6, pp. 721-731, https://doi.org/10.1103/PhysRev.41.721.

58. N. Inostroza, J. R. Letelier and M. L. Senent, "On the numerical determination of Dunham's coefficients: An application to $X^1\Sigma^+$HCl isotopomers", *J. Mol. Struct.: THEOCHEM*, 2010, **947**, 1, pp. 40-44, https://doi.org/10.1016/j.theochem.2010.01.037.

59. C. P. Rinsland, M. A. H. Smith, A. Goldman, V. M. Devi and D. C. Benner, "The Fundamental Bands of $H^{35}Cl$ and $H^{37}Cl$: Line Positions from High-Resolution Laboratory Data", *J. Mol. Spectrosc.*, 1993, **159**, 1, pp. 274-278, https://doi.org/10.1006/jmsp.1993.1124.

60. P. Maslowski, K. F. Lee, A. C. Johansson, A. Khodabakhsh, G. Kowzan, L. Rutkowski, A. A. Mills, C. Mohr, J. Jiang, M. E. Fermann and A. Foltynowicz, "Surpassing the path-limited resolution of Fourier-transform spectrometry with frequency combs", *Phys. Rev. A*, 2016, **93**, 2, 021802, https://doi.org/10.1103/PhysRevA.93.021802.

61. A. S. Pine, A. Fried and J. W. Elkins, "Spectral intensities in the fundamental bands of HF and HCl", *J. Mol. Spectrosc.*, 1985, **109**, 1, pp. 30-45, https://doi.org/10.1016/0022-2852(85)90049-9.

62. M. De Rosa, C. Nardini, C. Piccolo, C. Corsi and F. D'Amato, "Pressure broadening and shift of transitions of the first overtone of HCl", *Appl. Phys. B: Lasers Opt.*, 2001, **72**, 2, pp. 245-248, https://doi.org/10.1007/s003400000449.

63. T. Tomberg, M. Vainio, T. Hieta and L. Halonen, "Sub-parts-per-trillion level sensitivity in trace gas detection by cantilever-enhanced photo-acoustic spectroscopy", *Sci. Rep.*, 2018, **8**, 1, 1848, https://doi.org/10.1038/s41598-018-20087-9.

Supplementary information

# Infrared spectroscopy of radioactive hydrogen chloride H$^{36}$Cl

Santeri Larnimaa[a],*, Markku Vainio[a,b], Ville Ulvila[c]

[a] Department of Chemistry, University of Helsinki, Helsinki, Finland

[b] Photonics Laboratory, Physics Unit, Tampere University, Tampere, Finland

[c] VTT Technical Research Centre of Finland Limited, Espoo, Finland

* Corresponding author

      Postal address: Department of Chemistry, University of Helsinki, P.O. Box 55, FI-00014 Helsinki, Finland

      E-mail address: santeri.larnimaa@helsinki.fi

This document is organized as follows:

Supplementary Note 1: The use of Lorentzian functions to model the absorption peaks is justified.

Supplementary Note 2: The linewidths and integrated absorption coefficients (that reveal the relative intensities) of the H$^{36}$Cl fundamental rovibrational band, including an uncertainty analysis, are reported.

Supplementary Note 3: The H$^{36}$Cl linewidths and relative intensities to those of the stable isotopologues H$^{35}$Cl and H$^{37}$Cl are compared.

Supplementary Note 4: Our estimate of the gaseous H$^{36}$Cl content in the sample is explained.

# Supplementary Note 1: Justification of using Lorentzian functions to model the absorption peaks

Based on the HITRAN data for the stable isotopologues H$^{35}$Cl and H$^{37}$Cl [1,2], the expected values for the pressure-broadened HWHMs of the fundamental band lines P(10)–R(10) are larger than 0.013 cm$^{-1}$ (0.4 GHz), less than 0.090 cm$^{-1}$ (2.7 GHz) and 0.048 cm$^{-1}$ (1.4 GHz) on average. Although these expected HWHMs include the self-broadenings [3], the effect is generally small (less than 0.0005 cm$^{-1}$ or 15 MHz for all lines) due to the relatively low total amount of HCl (2.7 mbar). The theoretical Doppler HWHMs are less than 0.0033 cm$^{-1}$ (0.1 GHz), whereas the HWHM of the assumed instrument line shape (ILS) function is 0.0095 cm$^{-1}$ (0.3 GHz). A rule of thumb states that instrumental broadening is negligible if the instrumental linewidth is less than one third of the true line shape [4]. Therefore, except for the narrowest lines far from the band center, pressure broadening dominates the linewidths. For this reason, we modeled the absorption peaks as Lorentzian functions and report the fitted linewidths as the pressure-broadened linewidths. However, we considered the Doppler and instrument broadening induced inaccuracies by combining $\Delta\gamma = \sqrt{\gamma_{\text{fit}}^2 - \gamma_{\text{D}}^2 - \gamma_{\text{ILS}}^2} - \gamma_{\text{fit}}$ with the fit uncertainty and other uncertainty sources as discussed in the uncertainty analysis (Supplementary Note 2). Here, $\gamma_{\text{fit}}$ is the fitted HWHM of the Lorentzian function, $\gamma_{\text{D}}$ is the Doppler HWHM, and $\gamma_{\text{ILS}}$ is the HWHM of the assumed ILS function. The estimate stems from the assumption that the total observed linewidth can be approximated by $\sqrt{\gamma_{\text{L}}^2 + \gamma_{\text{D}}^2 + \gamma_{\text{ILS}}^2}$, which, according to numerical simulations that we performed, works reasonably well with the Norton-Beer medium apodization function. The minimum and maximum values for $\Delta\gamma$ are 0.00053 cm$^{-1}$ (16 MHz) and 0.00363 cm$^{-1}$ (109 MHz), respectively, which indeed implies that the Doppler and instrumental broadening effects are generally small.

# Supplementary Note 2: Linewidths, integrated absorption coefficients, and uncertainty analysis

Supplementary Table 1 lists the H$^{36}$Cl fit results for the linewidths (HWHM pressure broadenings) and the integrated absorption coefficients (line areas). The numbers in parentheses are one standard deviation uncertainties in least significant digits as obtained from the fit. The total uncertainty columns contain the combined effect of the fit uncertainties and other uncertainty sources for the respective line parameters at 68 % confidence level. Supplementary Table 2 contains a breakdown of the typical magnitudes of the different uncertainty contributions.

For the HWHMs, the total uncertainty estimates stated in Supplementary Tables 1 and 2 consist of the fit uncertainty, effect of the baseline determination uncertainty, uncertainty in the measurement pressure, and the effect of the ILS function. The effect of self-broadenings and temperature are negligible. The uncertainty of the 100 % transmission baseline has been obtained via numerical simulations where we systematically applied variation to the baseline and observed how the fit results changed. The effect of pressure has been estimated assuming 1 % (ca. 10 mbar) pressure uncertainty and the fact that pressure broadening is proportional to pressure. For the strongest lines, all the uncertainty sources are important. For the weakest lines, the instrumental broadening dominates. The total uncertainties are 2–5 % for lines P(7)–R(6). Note that for these lines the rule of thumb $\gamma_{\text{ILS}} < 1/3\, \gamma_{\text{tot}}$ holds.

For the integrated absorption coefficients, the total uncertainty estimates stated in Supplementary Tables 1 and 2 consist of the fit uncertainty, the effect of the ILS function, the effect of the baseline uncertainty, uncertainty in the measurement temperature, and uncertainty in the absorption path length. The effect of the ILS function has been estimated based on tests where we simulated instrument broadened absorption peaks and observed how well a Lorentz fit recovered the original line parameters. The temperature effect has been estimated by calculating how a 2-K temperature change affects the assumed HITRAN H$^{35}$Cl line intensities according to the line intensity temperature dependence equation found in HITRAN documentation [5]. The absorption path length uncertainty stems from a possible tilt of the sample cell with respect to the light propagation direction. It is this path length uncertainty that has the largest contribution to the total uncertainty of the strongest lines. For the weakest lines, all uncertainty contributions are important. The total uncertainties are 2–5 % for lines P(8)–R(8).

Supplementary Table 1. Pressure-broadened linewidth $\gamma$ (HWHM) and integrated absorption coefficient $\alpha_{\text{int}}$ (line area) results for the H$^{36}$Cl fundamental band. The numbers in parentheses are one standard deviation fit uncertainties in least significant digits and the values in columns denoted with $\sigma$ are our estimates of the total uncertainties of the respective line parameters at 68 % confidence level. The symbol $\gamma_{\text{ref}}$ refers to the corresponding pressure-broadened HWHMs of the H$^{35}$Cl isotopologue as obtained from the HITRAN database. Similarly, $\alpha_{\text{ref}}$ refers to the expected integrated absorption coefficients (Supplementary Note 3) calculated using the HITRAN H$^{35}$Cl line intensities, and a number density corresponding to 1.60 mbar partial pressure, i.e., our estimate of the amount of H$^{36}$Cl in the sample (Supplementary Note 4). The final column in the table thus compares the relative intensities of H$^{36}$Cl to those of H$^{35}$Cl. Note that all the values in the table are given at 1 atm pressure and 296 K temperature.

| Line | $\gamma$ (cm$^{-1}$) | $\sigma\gamma$ (%) | $(\gamma - \gamma_{\text{ref}})/\gamma$ (%) | $\alpha_{\text{int}}$ (10$^{-2}$ cm$^{-2}$) | $\sigma\alpha_{\text{int}}$ (%) | $(\alpha_{\text{int}} - \alpha_{\text{ref}})/\alpha_{\text{int}}$ (%) |
|---|---|---|---|---|---|---|
| P(10) | 0.0237(15) | 15 | 32 | 0.067(3) | 13 | 15 |
| P(9) | 0.0228(7) | 12 | 10 | 0.142(4) | 6 | 2 |
| P(8) | 0.0293(5) | 7 | 9 | 0.299(4) | 4 | -1 |
| P(7) | 0.0373(4) | 5 | 9 | 0.561(4) | 3 | -4 |
| P(6) | 0.0495(3) | 3 | 10 | 0.978(4) | 2 | -2 |
| P(5) | 0.0603(3) | 2 | 10 | 1.434(5) | 2 | -4 |
| P(4) | 0.0730(3) | 2 | 11 | 1.932(5) | 2 | -1 |
| P(3) | 0.0792(3) | 2 | 8 | 2.133(6) | 2 | -1 |
| P(2) | 0.0861(3) | 2 | 7 | 1.929(6) | 2 | 1 |
| P(1) | 0.0938(6) | 2 | 5 | 1.181(5) | 2 | 3 |
| R(0) | 0.0920(6) | 2 | 2 | 1.256(6) | 2 | 2 |
| R(1) | 0.0863(3) | 2 | 8 | 2.184(6) | 2 | 1 |
| R(2) | 0.0800(2) | 2 | 10 | 2.594(5) | 2 | 0 |
| R(3) | 0.0732(2) | 2 | 14 | 2.488(5) | 2 | -1 |
| R(4) | 0.0597(2) | 2 | 9 | 2.029(5) | 2 | -1 |
| R(5) | 0.0477(2) | 3 | 9 | 1.445(5) | 2 | 0 |
| R(6) | 0.0373(3) | 4 | 11 | 0.920(5) | 2 | 2 |
| R(7) | 0.0288(3) | 7 | 10 | 0.516(4) | 3 | 4 |
| R(8) | 0.0235(4) | 11 | 14 | 0.261(4) | 5 | 7 |
| R(9) | 0.0202(7) | 15 | 21 | 0.120(3) | 7 | 11 |
| R(10) | 0.0156(12) | 27 | 15 | 0.046(3) | 13 | 8 |

Supplementary Table 2. A breakdown of the typical magnitudes of the different uncertainty contributions to the total uncertainties of the H$^{36}$Cl fundamental band linewidths and integrated absorption coefficients listed in Supplementary Table 1.

| Line parameter | Uncertainty type | Min–max | Median |
|---|---|---|---|
| Pressure-broadened linewidth $\gamma$ | Fit | 0.3–8 % | 0.6 % |
| | Baseline | 0.3–10 % | 0.9 % |
| | Pressure | 1 % (ca. 10 mbar) | 1 % (ca. 10 mbar) |
| | ILS | 0.6–24 % | 3 % |
| | Total | 2–27 % | 3 % |
| Integrated absorption coefficient (line area) $\alpha_{\text{int}}$ | Fit | 0.2–6 % | 0.5 % |
| | ILS | 0.1–5 % | 0.3 % |
| | Baseline | 0.3–12 % | 0.7 % |
| | Temperature | 0.1–4 % (2 K) | 0.7 % (2 K) |
| | Path length | 1 % (1 mm) | 1 % (1 mm) |
| | Total | 2–13 % | 2 % |

# Supplementary Note 3: Comparison of the $H^{36}Cl$ linewidths and relative intensities to those of the stable isotopologues $H^{35}Cl$ and $H^{37}Cl$

In the fourth column of Supplementary Table 1 (denoted by $(\gamma - \gamma_{\text{ref}})/\gamma$), we compare the measured HWHMs to the corresponding HITRAN values of $H^{35}Cl$. This comparison is interesting because there are multiple reports on $H^{35}Cl$ and $H^{37}Cl$ linewidths where the differences between the different isotopologues are small, typically within the stated measurement uncertainties that are of the order of a few percent [3,6-12]. However, the HWHMs of $H^{36}Cl$ in this study are systematically larger than the HITRAN values for $H^{35}Cl$. The median of the differences is 10 %, whereas the median of the estimated total uncertainties is 3 %. We also determined the HWHMs for the stable isotopologues $H^{35}Cl$ and $H^{37}Cl$ from our experimental data and these also seemed to follow this trend; the median of the differences between the experimental HWHMs for $H^{35}Cl$ ($H^{37}Cl$) and the corresponding HITRAN values is 8 % (7 %). The HITRAN data for the linewidths are based on laser spectroscopy measurements by Pine and Looney [2], who estimated 2–4 % total uncertainty for their results. On the other hand, Asfin et al. [7] reported in a recent study widths that were approximately 5 % smaller than the results of Pine and Looney. The reasons for the deviations of our results from the previously reported linewidth values are unknown. The determined HWHMs of the three HCl isotopologues together with the corresponding HITRAN data are plotted in Supplementary Figure 1.

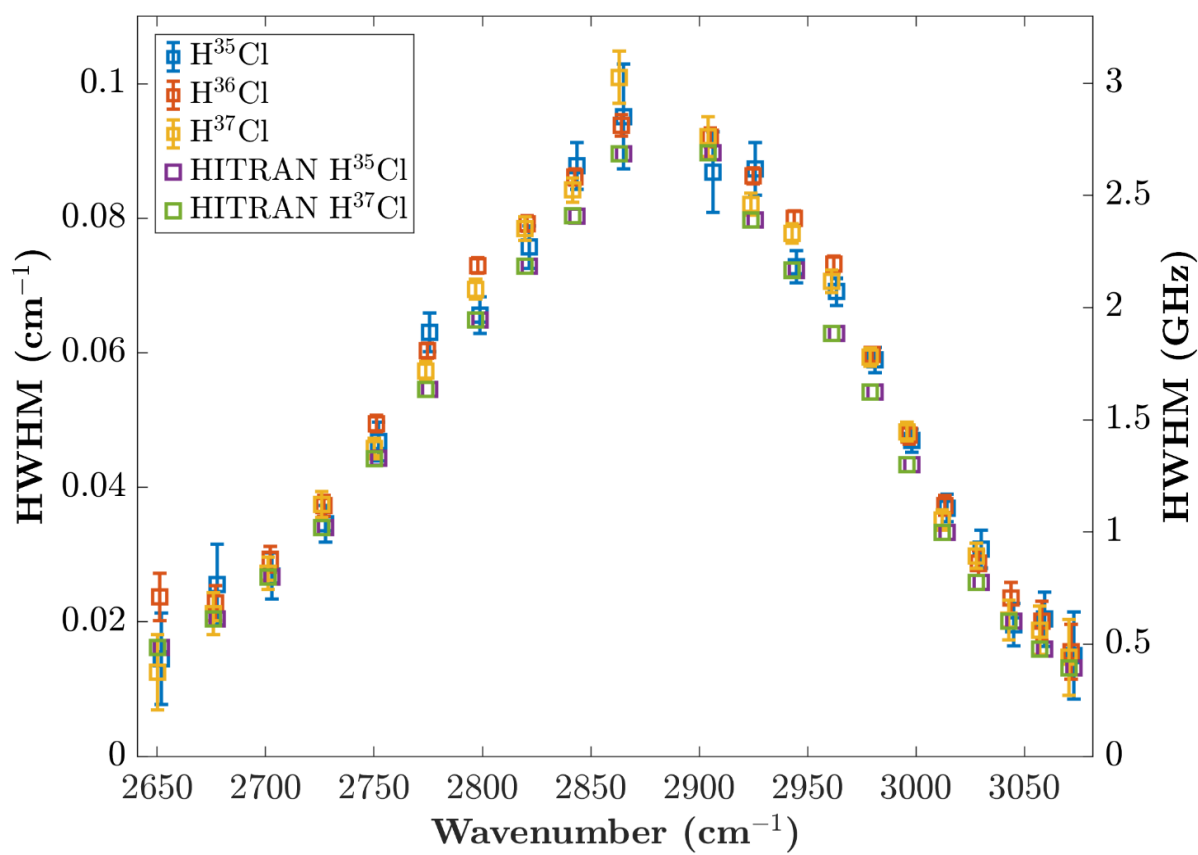

Supplementary Figure 1. Plot of the determined fundamental band linewidths (HWHM) of the three HCl isotopologues together with the corresponding HITRAN values for the stable isotopologues. Apart from self-broadenings, note that HITRAN assumes the $H^{35}Cl$ air broadenings also for $H^{37}Cl$. The error bars for the $H^{36}Cl$ linewidths correspond to the estimated total uncertainties in Supplementary Table 1. The error bars for the stable isotopologues were determined in a similar manner.

For the final column of Supplementary Table 1, we multiplied the H$^{35}$Cl HITRAN line intensities with a number density corresponding to 1.6 mbar of H$^{36}$Cl (see Supplementary Note 4) to obtain the expected integrated absorption coefficients $\alpha_{\text{ref}}$. *The column thus compares the relative intensities of H$^{36}$Cl to those of H$^{35}$Cl*. The agreement between the values is excellent, which is further illustrated in Supplementary Figure 2, where we plotted the measured integrated absorption coefficients divided by the aforementioned number densities together with the corresponding HITRAN line intensities for the stable isotopologues. The results imply that the line intensities of H$^{36}$Cl are indeed similar to those of the stable isotopologues. However, it is noteworthy that the line intensities depend slightly on the wavenumber of the transition and on the mass of the molecule [13,14].

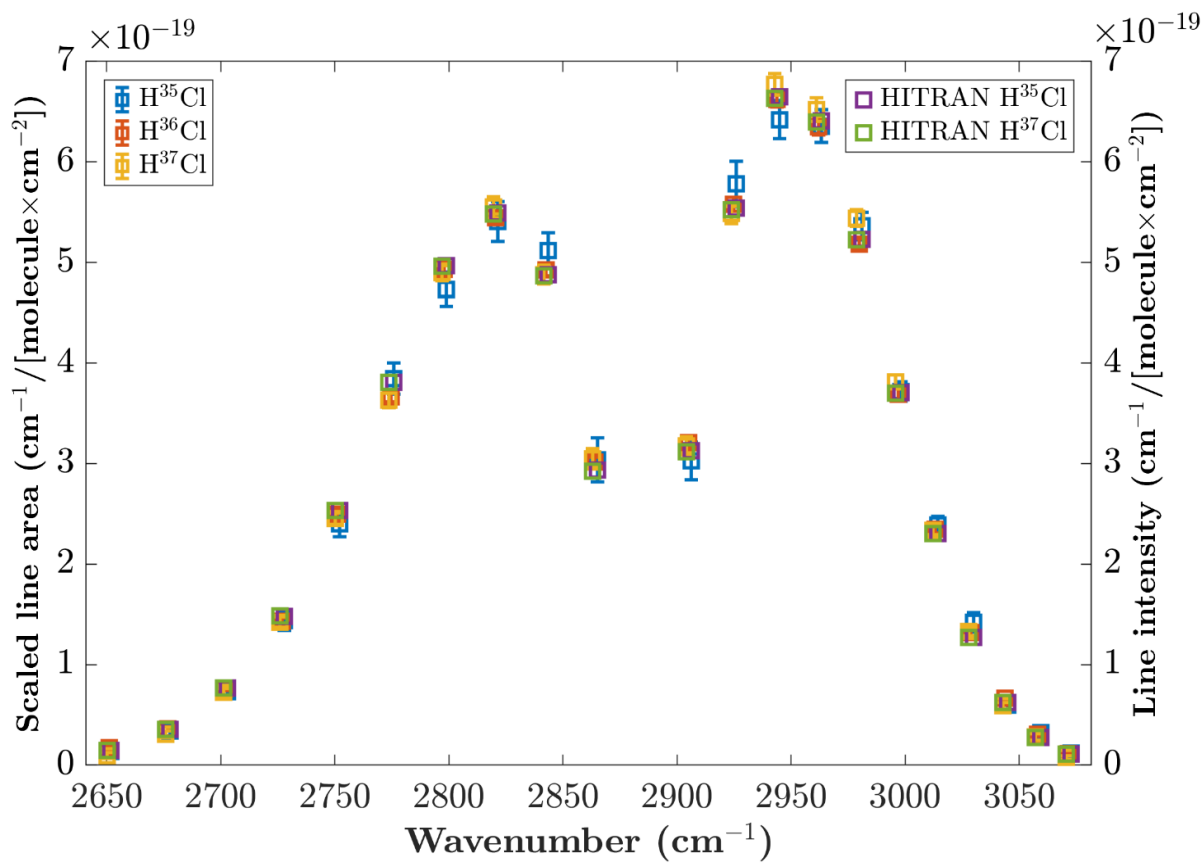

Supplementary Figure 2. Plot of the experimental integrated absorption coefficients, i.e., line areas (squares, left axis) of the three HCl isotopologues divided by the estimated number densities corresponding to 0.39 mbar, 1.60 mbar, and 0.76 mbar partial pressures for H$^{35}$Cl, H$^{36}$Cl, and H$^{37}$Cl, respectively, and of the HITRAN H$^{35}$Cl and H$^{37}$Cl line intensities (diamonds, right axis) divided by the corresponding natural abundances given in HITRAN. The error bars for H$^{36}$Cl correspond to the estimated total uncertainties in Supplementary Table 1. The error bars for the other two isotopologues have been determined similarly. The plot illustrates how the relative intensities of the measured lines behave compared to the HITRAN data of the stable isotopologues.

# Supplementary Note 4: Explanation for our estimate of the gaseous H$^{36}$Cl content in the sample

The integrated absorption coefficients $\alpha_\text{int}$ in the fifth column of Supplementary Table 1 are proportional to the H$^{36}$Cl line intensities and number density. Since we do not have accurate external knowledge of the H$^{36}$Cl number density, we cannot determine the absolute line intensities from our integrated absorption coefficient data. The integrated absorption coefficients thus reveal only the relative intensities of the lines. However, similar to the linewidths, there are several prior reports on H$^{35}$Cl and H$^{37}$Cl where the differences between the intensities of the different isotopologues are small and typically within the stated measurement uncertainties [6,11,12,14]. Therefore, we can estimate the number density of H$^{36}$Cl in the sample if we assume the HITRAN H$^{35}$Cl line intensities. We performed this by calculating the weighted average of the ratio of the H$^{36}$Cl integrated absorption coefficients to the HITRAN H$^{35}$Cl line intensities [1,15]. The weights given to the data points are the inverses of the respective fit uncertainties. The obtained result is a number density corresponding to 1.60 mbar partial pressure, which is our estimate of the amount of H$^{36}$Cl in the sample. A similarly weighted standard deviation is 0.04 mbar. The same calculation yields 0.39(2) mbar and 0.76(4) mbar for H$^{35}$Cl and H$^{37}$Cl, respectively. Note that the HITRAN line intensities include the natural abundances of the different isotopologues, which we considered in the calculations.

# Supplementary References

1. I. E. Gordon, L. S. Rothman, C. Hill, R. V. Kochanov, Y. Tan, P. F. Bernath, M. Birk, V. Boudon, A. Campargue, K. V. Chance, B. J. Drouin, J.-M. Flaud, R. R. Gamache, J. T. Hodges, D. Jacquemart, V. I. Perevalov, A. Perrin, K. P. Shine, M.-A. H. Smith, J. Tennyson, G. C. Toon, H. Tran, V. G. Tyuterev, A. Barbe, A. G. Császár, V. M. Devi, T. Furtenbacher, J. J. Harrison, J.-M. Hartmann, A. Jolly, T. J. Johnson, T. Karman, I. Kleiner, A. A. Kyuberis, J. Loos, O. M. Lyulin, S. T. Massie, S. N. Mikhailenko, N. Moazzen-Ahmadi, H. S. P. Müller, O. V. Naumenko, A. V. Nikitin, O. L. Polyansky, M. Rey, M. Rotger, S. W. Sharpe, K. Sung, E. Starikova, S. A. Tashkun, J. V. Auwera, G. Wagner, J. Wilzewski, P. Wcisło, S. Yu and E. J. Zak, "The HITRAN2016 molecular spectroscopic database", *J. Quant. Spectrosc. Radiat. Transfer*, 2017, **203**, pp. 3-69, https://doi.org/10.1016/j.jqsrt.2017.06.038.

2. A. S. Pine and J. P. Looney, "N$_2$ and air broadening in the fundamental bands of HF and HCl", *J. Mol. Spectrosc.*, 1987, **122**, 1, pp. 41-55, https://doi.org/10.1016/0022-2852(87)90217-7.

3. A. S. Pine and A. Fried, "Self-broadening in the fundamental bands of HF and HCl", *J. Mol. Spectrosc.*, 1985, **114**, 1, pp. 148-162, https://doi.org/10.1016/0022-2852(85)90344-3.

4. A. Sieghard, K. A. Keppler and M. Quack, *High-resolution Fourier Transform Infrared Spectroscopy*, *Handbook of High-resolution Spectroscopy*, Vol. 2, eds. M. Quack and F. Merkt, John Wiley & Sons, 2011.

5. M. Šimečková, D. Jacquemart, L. S. Rothman, R. R. Gamache and A. Goldman, "Einstein A-coefficients and statistical weights for molecular absorption transitions in the HITRAN database", *J. Quant. Spectrosc. Radiat. Transfer*, 2006, **98**, 1, pp. 130-155, https://doi.org/10.1016/j.jqsrt.2005.07.003.

6. G. Li, A. Serdyukov, M. Gisi, O. Werhahn and V. Ebert, "FTIR-based measurements of self-broadening and self-shift coefficients as well as line strength in the first overtone band of HCl at 1.76µM", *J. Quant. Spectrosc. Radiat. Transfer*, 2015, **165**, pp. 76-87, https://doi.org/10.1016/j.jqsrt.2015.06.021.

7. R. E. Asfin, A. V. Domanskaya and C. Maul, "Broadening and shifting coefficients of rotation–vibrational lines in the fundamental and first overtone bands of HCl and HBr induced by oxygen and air", *J. Quant. Spectrosc. Radiat. Transfer*, 2013, **130**, pp. 296-303, https://doi.org/10.1016/j.jqsrt.2013.07.014.

8. R. E. Asfin, A. V. Domanskaya, C. Maul and M. O. Bulanin, "Nitrogen-induced broadening and shift coefficients of rotation–vibrational lines in the fundamental and first overtone bands of HCl and HBr", *J. Mol. Spectrosc.*, 2012, **282**, pp. 9-13, https://doi.org/10.1016/j.jms.2012.10.009.

9. M. Tudorie, T. Földes, A. C. Vandaele and J. Vander Auwera, "$CO_2$ pressure broadening and shift coefficients for the 1–0 band of HCl and DCl", *J. Quant. Spectrosc. Radiat. Transfer*, 2012, **113**, 11, pp. 1092-1101, https://doi.org/10.1016/j.jqsrt.2012.01.025.

10. M. De Rosa, C. Nardini, C. Piccolo, C. Corsi and F. D'Amato, "Pressure broadening and shift of transitions of the first overtone of HCl", *Appl. Phys. B: Lasers Opt.*, 2001, **72**, 2, pp. 245-248, https://doi.org/10.1007/s003400000449.

11. C. L. Lin, E. Niple, J. H. Shaw, W. M. Uselman and J. G. Calvert, "Line parameters of HCl obtained by simultaneous analysis of spectra", *J. Quant. Spectrosc. Radiat. Transfer*, 1978, **20**, 6, pp. 581-591, https://doi.org/10.1016/0022-4073(78)90029-8.

12. R. A. Toth, R. H. Hunt and E. K. Plyler, "Line strengths, line widths, and dipole moment function for HCl", *J. Mol. Spectrosc.*, 1970, **35**, 1, pp. 110-126, https://doi.org/10.1016/0022-2852(70)90169-4.

13. R. Herman and R. F. Wallis, "Influence of Vibration-Rotation Interaction on Line Intensities in Vibration-Rotation Bands of Diatomic Molecules", *J. Chem. Phys.*, 1955, **23**, 4, pp. 637-646, https://doi.org/10.1063/1.1742069.

14. A. S. Pine, A. Fried and J. W. Elkins, "Spectral intensities in the fundamental bands of HF and HCl", *J. Mol. Spectrosc.*, 1985, **109**, 1, pp. 30-45, https://doi.org/10.1016/0022-2852(85)90049-9.

15. G. Li, I. E. Gordon, P. F. Bernath and L. S. Rothman, "Direct fit of experimental ro-vibrational intensities to the dipole moment function: Application to HCl", *J. Quant. Spectrosc. Radiat. Transfer*, 2011, **112**, 10, pp. 1543-1550, https://doi.org/10.1016/j.jqsrt.2011.03.014.